\documentclass[journal]{vgtc}                          

\ifpdf
  \pdfoutput=1\relax                   
  \usepackage{graphicx}                
  \DeclareGraphicsExtensions{.pdf,.png,.jpg,.jpeg} 
\else
  \usepackage{graphicx}                
  \DeclareGraphicsExtensions{.eps}     
\fi%

\graphicspath{{figures_new/}{pictures/}{images/}{./}} 

\usepackage{microtype}                 
\PassOptionsToPackage{warn}{textcomp}  
\usepackage{textcomp}                  
\usepackage{mathptmx}                  
\usepackage{times}                     
\usepackage{cite}                      
\usepackage{tabu}                      
\usepackage{booktabs}                  

\usepackage{todonotes}
\usepackage[caption=false, font=footnotesize]{subfig}
\usepackage{multirow, array}
\usepackage{graphicx}
\usepackage{comment}
\usepackage{float}
\usepackage{balance}

\newcommand{\review}[1]{{\textcolor{black}{#1}}}
\newcommand{\reviewnew}[1]{{\textcolor{black}{#1}}}

\onlineid{1166}

\vgtccategory{Research}

\vgtcinsertpkg

\vgtcpapertype{application/design study}

\teaser{
	\centering
	\includegraphics[width=\linewidth]{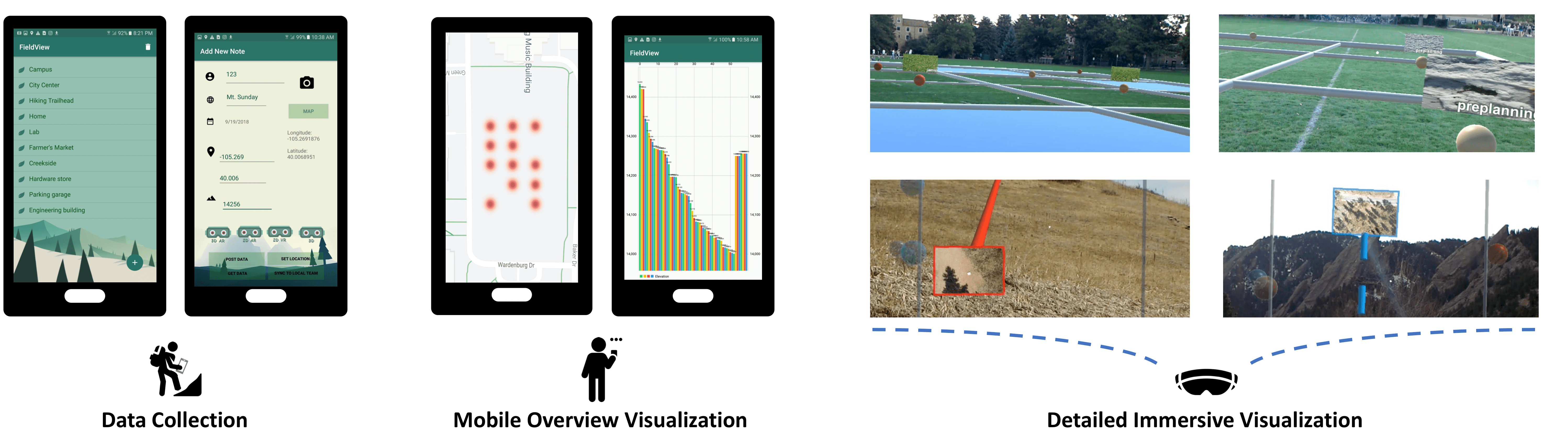}
	\caption{We explore how visualization can enhance data use in the field for domains such as earth science and emergency response. 
		\review{Working with field analysts, we elicit}
		recommendations for how \review{mobile and immersive }technologies might bridge spatial and temporal gaps in data collection and analysis. We use these recommendations to develop FieldView, an extensible prototype data collection and visualization system that uses mobile 
		overviews
		and situated AR visualizations to communicate
		data about 
		\review{on-going operations}.}	\label{fig:teaser}
}

\title{Designing for Mobile and Immersive Visual Analytics in the Field}

\author{Matt Whitlock, Keke Wu, Danielle Albers Szafir}
\authorfooter{
	\item
	Matt Whitlock is with the University of Colorado Boulder. E-mail: matthew.whitlock@colorado.edu.
	\item
	Keke Wu is with the University of Colorado Boulder. E-mail: keke.wu@colorado.edu.
	\item
	Danielle Albers Szafir is with the University of Colorado Boulder. E-mail: danielle.szafir@colorado.edu.
}

\abstract{
	Data collection and analysis in the field is critical for operations in domains such as environmental science and public safety. 
	However, field workers currently face data- and platform-oriented issues in efficient data collection and analysis in the field, such as limited connectivity, screen space, and attentional resources.
	In this paper, we explore how visual analytics tools might transform field practices by more deeply integrating data into these operations. 
	We use a design probe coupling mobile, cloud, and immersive analytics components to guide interviews with ten experts from five domains to explore how visual analytics could support data collection and analysis needs in the field. 
	\review{
		The results identify shortcomings of current approaches and target scenarios and design considerations for future field analysis systems.}
	We embody these findings in \review{FieldView, an extensible, open-source}  prototype designed to support critical use cases for situated field analysis. 
	Our findings suggest the potential for integrating mobile and immersive technologies to enhance data's utility for various field operations and 
	new directions for visual analytics tools to transform fieldwork. 
}

\CCScatlist{
  \CCScatTwelve{Human-centered computing}{Visu\-al\-iza\-tion}{Visu\-al\-iza\-tion techniques}{Treemaps};
  \CCScatTwelve{Human-centered computing}{Visu\-al\-iza\-tion}{Visualization design and evaluation methods}{}
}

\begin{document}

\maketitle



\section{Introduction}

Data-oriented decision making is transforming practices in a broad variety of domains. 
Applications in earth science \review{\cite{eeckhaut2007use}}, geology \review{\cite{pavlis2010computer}}, and emergency response \review{\cite{palen2016crisis}} all leverage data collected in the field to describe the state of complex environments.
Field analysts collect data to model the changing state of field sites and share information across teams to increase situational awareness and deepen scientific and operational understanding.
However, \review{our interviews with field workers across multiple domains reveal that} current practices for working with data in the field 
rely heavily on decoupled solutions that separate data analysis from the spatial and temporal contexts that data describes. 
Data is collected in the field but analyzed in operations centers or remote laboratories.
\review{Analysts report that this} separation limits \review{their} abilities to use this data in on-going operations: remote analysts lack the context and local situational awareness of people in the field; people in the field lack tools to access, analyze, and act on data while in the field. 
This work explores challenges and opportunities for using visualization to support data analysis in the field. 

\begin{figure}[t]
	\centering
	\includegraphics[width=\columnwidth]{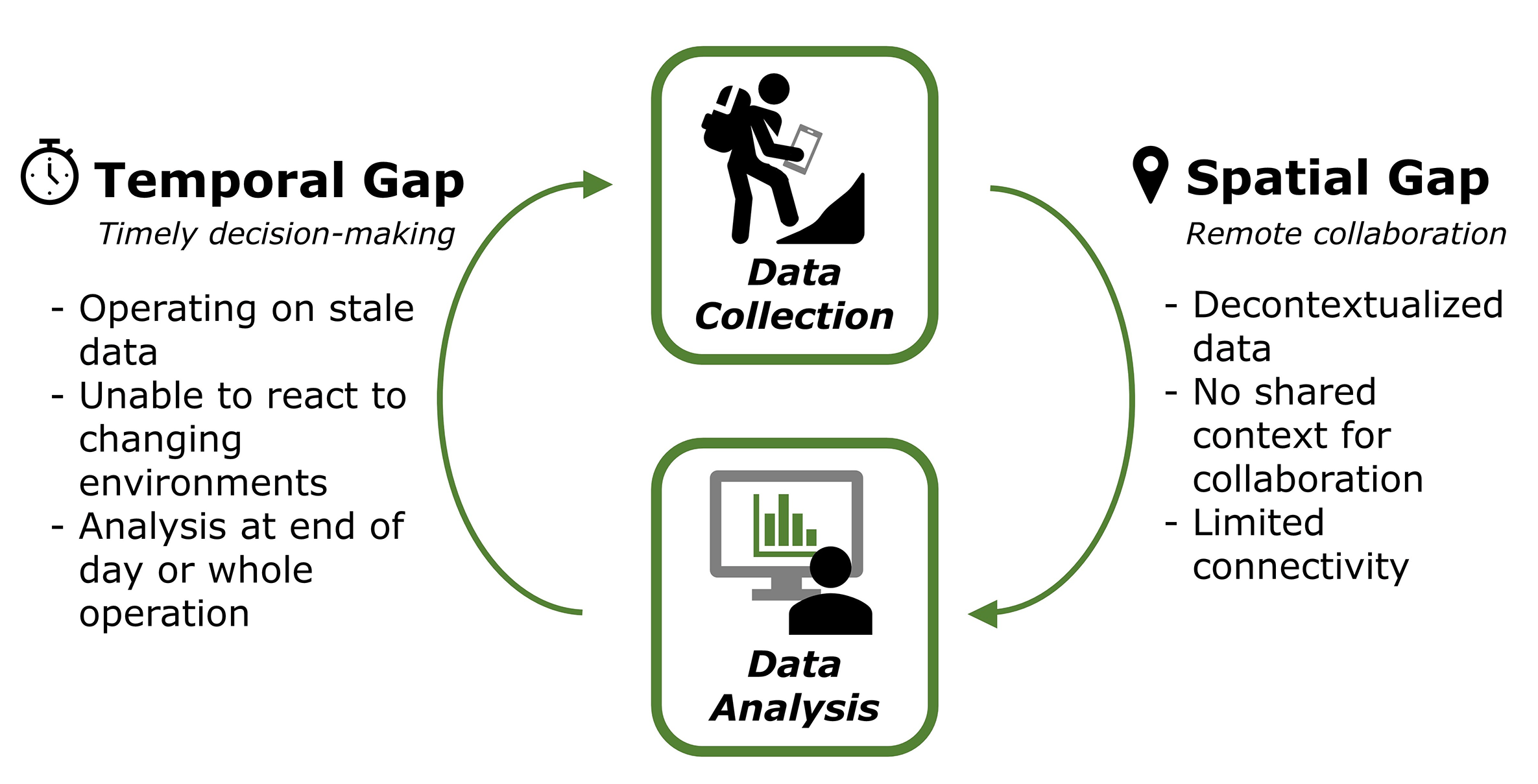}
	\caption{Current field practices separate data collection and analysis, leading to temporal (separating data collection and analysis by time) and spatial (decontextualizing data from the environment) gaps that 
		hinder effective analysis and decision making with field data. 
	}
	\label{fig:SpatialTemporal}
	\vspace{-5px}
\end{figure}

The lack of support for analytics in the field means that decision making typically occurs in remote operations centers either during or after operations.
Current practices for fieldwork require analysts to first preplan their operations based on data from previous collection efforts and archival data streams. Analysts then collect new data either on mobile devices or in field notebooks, physically transport data to a central location to synchronize
with other sources, and replan subsequent collection efforts and operational practices based on the newly revised data \review{\cite{cutter2003gi,pavlis2010computer}}.
This workflow decontextualizes collected data from the environment, prevents analysts from reacting to new data, and obscures possible errors in data collection by creating spatial and temporal gaps between data collection and analysis (Fig. \ref{fig:SpatialTemporal}).
Spatial gaps in understanding arise when remote analysts lack the physical context 
surrounding field data, resulting in a lack of visual grounding for remote analysis and 
limited shared context during real-time operations. 
Temporal gaps stem from an inability to analyze data during an operation, resulting in stale data being used in dynamic environments and an inability to make timely decisions.
Temporal gaps could jeopardize a day's work or even the entire operation due to data quality issues and reduced global awareness of the state of the broader operation. \review{While domain experts noted that these gaps are acceptable for some complex analyses (e.g., those requiring heavy computation or in-depth exploration), improving access to field data could transform many aspects of field practices by expanding field analysts' situational awareness.}


We explore how mobile and immersive analytics tools can \review{begin to} bridge spatial and temporal gaps between data collection and analysis in fieldwork. Specifically,
we examine how coupling these technologies might increase contextual and situational awareness in field practices 
to improve data collection, sharing, analysis, and decision making by better connecting field analysts with their data. 
Mobile technologies offer portable connectivity to access and synchronize data in the field and can offer at-a-glance overviews that help summarize \review{an operation's} current state
\cite{brehmer2019visualizing}.
However, mobile technologies suffer from several limitations 
\review{in the field.} 
For example, interacting with mobile devices is challenging while wearing gloves and forces analysts to divide their attention between the visualization and environment. Immersive visualization techniques can resolve these issues by visually embedding data in the physical environment \cite{willett2017embedded}; however, we have limited knowledge of how to design effective AR visualizations. Our goal in this paper is to elicit insight into how visualization tools can be designed to effectively support the needs of field data analysis to provide preliminary steps towards a deeper integration of data into field operations. 

This paper presents an exploratory investigation into how visualization tools could support increased situated awareness and optimize data-driven decision making in the field.
Building on preliminary conversations with field workers in domains ranging from earth science to emergency response, we created an interactive design probe architected to support an integrated data collection and visual analysis workflow
\review{combining} mobile and immersive tools.
We use this probe to elicit formal recommendations for visual analytics tools for field work and develop a preliminary prototype based on these recommendations.
The prototype 
\review{addresses} three target scenarios identified during the interviews where a lack of access to field data causes significant operational challenges: \review{team coordination, data quality validation, and integrating autonomous sensors}. Our results suggest new opportunities for leveraging visualization to empower field operations through data. 

\textbf{Contributions: }
\review{Our primary contribution is a formative qualitative study used to establish preliminary design considerations, constraints, and scenarios for integrated data collection and analysis tools in the field. }
We 
worked with scientists and public safety officials to identify challenges in current field data practices
\review{and} 
develop a design probe demonstrating how mobile and immersive technologies might overcome these limitations.
This probe scaffolded interviews with 10 domain experts about the potential of complimentary mobile and immersive visualizations to transform field operations.
After synthesizing these interviews, we refined the workflow introduced in the design probe to develop FieldView, an open-source prototype system to address 
\review{three critical use scenarios identified by analysts (teaming, data quality validation, and data fusion;} Fig. \ref{fig:teaser}).
\section{Related Work}
\label{sec:RelatedWork}
Bridging the gap between data collection and analysis for field work requires
understanding how contextual awareness can inform analysis 
and the technologies and approaches that might enable such solutions. 
We focus on developing methods that provide field analysts ready access to important information to improve the quality of their data collection practices, interpretation, and operational decision making.
We build on 
work in situational awareness and contextual computing in HCI as well as mobile and immersive analytics 
to support these goals. 

\subsection{Contextual Awareness}
Current practices for analyzing field data limit contextual and situational awareness (SA). Technologies may consider many forms of SA. For example, Endesley et al. models three hierarchical phases of SA in dynamic systems \cite{endsley1995toward}: (1) perceiving elements in the environment, (2) comprehending the active situation based on disjoint elements, and (3) projecting the future status of a situation based on the status and dynamics of the current situation.
Systems can increase SA across these phases by designing for change detection, preparedness for interruption, goal reorientation, and detection of missed changes \cite{john2008staying}.

Increasing SA is especially critical in technologies for highly dynamic situations where new data frequently changes operational strategies, such as search-and-rescue and emergency response.
For example, Cao et al. 
found that people trusted their own situated understanding over the advice of agencies in predicting wildfire spread \cite{cao2017smoke}.
Kim et al. increase field responder SA by visualizing location-based data on mobile devices, allowing users to ``see through the fog'' \cite{kim2008mobile}. 
While these solutions increase awareness of areas invisible to users, they require divided attention and map complex data to a limited visual space.
We address these limitations by exploring how AR may supplement existing mobile data collection and visualization approaches, bypassing the need for device-to-environment context shifts by situating data directly within the physical environment to increase contextual awareness.

We ground our exploration of 
\review{SA} in visualization systems. These systems allow analysts to bring domain expertise and contextual awareness to data exploration and decision making by allowing analysts rather than algorithms to synthesize insights from available data.
For example, Chan et al. uses multiple spatially aware displays to increase SA in a command center \cite{Chan2016}.
Visual analytics systems may support 
\review{field work} by reducing reliance on verbal/radio communication, integrating information from multiple sources, and combining both streaming and manual data entry \cite{goubran2016modelling} as well as improving collaborative teaming \cite{kulyk2008situational}.
These technologies aim to provide domain experts 
with sufficient knowledge about the global state of the operational environment to make data-informed decisions in dynamic environments.
However, 
these solutions generally target a remote operator with a global view of the task at hand but limited understanding of the context of field data.
We instead aim to put these visualizations in the field, considering how mobile technologies and immersive visualization can support field analysis needs 
while mitigating limiting factors of current mobile approaches like divided attention and limited \review{screen} size.

\subsection{Mobile Data Collection and Visualization} 
\label{sec:MobileRelatedWork}
Mobile devices support portable data collection and sensing across a 
variety of applications \cite{ganti2011mobile,reades2007cellular,lane2010survey,Parikh2006Mobile,choe2015sleeptight,jo2017touchpivot,chen2019marvist}. 
Using mobile devices for data entry can enhance data quality over traditional physical map-based methods \cite{ryan1999enhanced, fletcher2003handheld}.
For example, Pascoe et al. discuss user interfaces for field data collection that
minimize the attention needed to use a device while maximizing benefits of contextual awareness \cite{pascoe2000using}. 
Tomlinson et al. 
show how mobile devices can support distributed data collection efforts \cite{tomlinson2009use}.
ESCAPE provides a middle-ware 
for exchanging information between mobile devices in emergency response \cite{truong2007escape}.
These studies show that simplicity and contextual awareness are critical components of successful field data collection. 

However, analyzing data on mobile devices can be difficult due to the limited display, storage, and computation capabilities of these technologies \cite{chittaro2006visualizing, langner2018v, melo2017hdr}. These challenges are exacerbated in field scenarios and extended to include other challenges such as decision time and privacy concerns \cite{flentge2008designing}. 
Solutions like Siren leverage context-aware paradigms to support peer-to-peer messaging based on environmental triggers, choosing to reduce cognitive load through automated analysis based on contextual data \cite{jiang2004siren}.
Though mobile technologies have limited display size and resolution, they readily support location-based visual analysis, including mobile tourist guides \cite{burigat2005location}, overview-detail map visualizations \cite{karstens2005presenting} and mobile device-mediated navigation tools \cite{burigat2007geographical, rakkolainen20013d}. These visualizations generally focus on a few critical data attributes 
to increase analysts' abilities to get important information at a glance \cite{wiehr2013drivesense}. 
While these systems provide useful at-a-glance overviews for general analysis or detail views for well-defined tasks, they use a small form factor that divides both data and analyst attention between the screen and the operational environment, limiting contextual integration into analysis and decision making. 
We build on these approaches to support location-based data collection and overview analysis and extend these approaches with immersive AR visualizations to enable deeper engagement with data. These AR visualizations blend data and context to more closely couple data exploration with the operational environment for applications ranging from forest ecology to public safety.

\subsection{Immersive Visualization for Fieldwork}
Immersive Analytics (IA) uses embodied data analysis for data understanding and decision making, generally through 3D virtual or augmented reality.
IA offers opportunities that standard 2D displays do not, including benefits associated with situated analytics, embodied data exploration, collaboration, and increased engagement \cite{marriott2018immersive}. 
\review{Prior work in IA} focuses on the benefits of situated analytics, especially how the use of augmented reality (AR) can enable contextually-aware data analysis by outfitting the physical environment with virtual data
(see Schmalsteig \& H\"ollerer for a survey \cite{schmalstieg2016augmented}).
These techniques can leverage data 
from disparate environments for situated analysis \cite{elmqvist2013ubiquitous}.
Such techniques include mid-air displays for map navigation \cite{dancu2015map} 
and cross-device displays for collaborative visualization \cite{badam2017visfer}.
Systems have used AR to directly annotate the environment with data for applications in construction and architectural oversight \cite{irizarry2013infospot}, energy aware smart homes \cite{jahn2010energy}, manufacturing \cite{Lukosch2015Providing,bell2002annotated}, and situated learning \cite{Santos2016,vcopivc2016playing}.
Other systems use AR to highlight critical information along 3D surfaces including rock folds \cite{gazconimmersive}, mammograms \cite{douglas2017augmented}, and physical bodies \cite{hoang2017augmented}. 
In our target domain, visualization overlays on proximal referents can help highlight previously searched paths in search-and-rescue \cite{willett2017embedded}.
These techniques offer interactive methods for engaging with spatial data that may outperform other mobile technologies such as tablets \cite{bach2018hologram}.

In fieldwork, these technologies may help resolve the ``field map shuffle'' where researchers constantly reference several potentially out-of-date sources for geographical 
\review{data} \cite{pavlis2010computer}.
For example, McCaffrey et al. 
theorize how portable stereo rendering might enable field-based visualization for complex geospatial models \cite{mccaffrey2005unlocking}. Pavlis et al. extend this vision to suggest how immersive visualization may improve map synthesis \cite{pavlis2010computer}. 
However, these efforts offer purely theoretical insight into such designs. 
More recent works have 
begun to explore preliminary designs for targeted analysis problems in field research.
For example, Gazc{\'o}n et al. 
allow field researchers to augment an existing view of mountains with lines highlighting geological folding patterns \cite{gazconimmersive}.
Ramakrishnaraja et al. 
visualize sensor data in AR to increase SA for oil and gas workers \cite{ramakrishnaraja2017unaware}.
However, these systems focus on a narrow set of well-defined analysis tasks and do not consider how field analysts can integrate incoming information. We extend these ideas to enable holistic situated analysis for field domains. We do so by first systematically understanding how immersive situated analysis in tandem with mobile devices
may better support data-oriented fieldwork.


\section{Preliminary Requirements Analysis}
\label{requirements}
\review{We aim to understand how visual analytics can improve field operations by bridging spatial and temporal gaps in field data usage.}
To characterize the problem space, we enumerate the anticipated requirements of \review{field VA systems} and identify limitations in existing approaches 
\review{in} preliminary \review{unstructured} interviews with four field analysts: one in public safety, two in wildland fire, and one in geological science.
\review{We synthesized challenges and requirements noted by all participants to derive four preliminary design considerations. Our design probe embodies these design considerations to elicit insight into how visual analytics can enhance field practices.} \reviewnew{While we found no studies specific to field data needs, several points raised in the interviews exemplify general data and technological challenges in emergency response. }

In these interviews, analysts \review{universally} identified \reviewnew{four} primary limitations that prevent them from adequately leveraging data in 
\review{the field:} divided attention, small form-factors, \reviewnew{touchscreen input,} and reliance on operations centers. \reviewnew{The first two factors echo concerns for mobile devices in GIS work \cite{burigat2007geographical}.}
Analysts currently preplan daily goals based on archival data from prior operations before entering the field, using mobile devices and paper to collect new data in the field.
\review{Due to limited support for data collection in the field, experts' current data consists of digital images collected during pre-op aerial fly-overs (wildland fire and public safety), archival data printed onto paper maps (e.g., topography, structural locations, and prior survey data; all domains), simple quantitative measurements from portable sensors streamed to a portable drive (e.g., canopy density, soil acidity; geological science and wildland fire), and hand-written qualitative observations (geological science). Analysts reported only having preplanning data and handwritten observations while on a mission and had no methods for engaging in any analysis tasks in the field.}
They \review{instead} synchronize and analyze \review{new} data after returning to the operations center or at a remote site after leaving the field, \reviewnew{forcing operations centers to resolve data collection inconsistencies \cite{Chan2016}}.
Current mobile-only solutions offer insights into how data could inform field practices, but require analysts to use notes and memories to integrate environmental context back into their analyses. 
Mobile visualizations offer a potential alternative, but 
\review{divide attention} between the display and the operational environment, introducing hazards similar to texting and driving \cite{lamble1999cognitive}. 

These limitations lead to heavier reliance on external analysis at an operations center or field site, but limited connectivity can cause information loss and substantial delays in data exchange. 
Operations centers also lack the contextual awareness of field analysts \reviewnew{as noted in Flentge et. al. \cite{flentge2008designing}}, 
while field analysts lack the awareness of multiple fieldsites granted to remote operators \reviewnew{as noted in Kim et. al. \cite{kim2008mobile}}. 
\review{As a result,} field analysts \review{enumerated three options for using data in current practices:} operate agnostic of field collected data, use stale data from daily preplanning, or rely on strictly verbal communications from the operations center. 
\review{All analysts felt that visual analytics tools could transform these options by increasing situational awareness in the field. The discussions made clear that field practices offer a rich design space that differs from conventional visual analytics tools. However, analysts had a limited sense of the capabilities of current devices for situated analysis and had difficulty reasoning about how these technologies might best augment their workflows. Our interviews elicited four design considerations for field analysis agreed upon by all participants: }  

\vspace{3pt}
\noindent
\textbf{R1--Offline \& Distributed Data Collection \review{\& Analysis}:} Field analysts often work in remote locations with limited to no connectivity\review{, inhibiting fully cloud-based solutions} \reviewnew{\cite{flentge2008designing}}. 
Most field analysts collect and log data in the field, store that the data on portable thumbdrives, and fuse that data with data from other efforts back at the remote operations center or lab. 
\review{Analysts noted that variations in data collection practices used by different teams can lead to format inconsistencies that further inhibit data sharing between teams.} \review{Aggregating and} visualizing new data in real time would improve decision making for team operations when connected and provide timely perspectives on their current situations when offline. 

\noindent
\textbf{R2--Merge Environmental Context \& Analysis:} Data collection in fieldwork attempts to capture important aspects of a fieldsite to build a broader understanding of the changing environments, functions, and scenarios present in the world \reviewnew{\cite{cutter2003gi}}. 
However, not all aspects of an environment can be quantified and captured in a database. 
Therefore, building effective tools for situated field analysis requires support for analysts to assess incoming data in tandem with its environmental context to increase contextual awareness and enrich decision making.

\noindent
\textbf{R3--Mitigate Information Overload:} Analysts often multitask in the field. They therefore need to get a quick sense of how their data fits into the current context without having to spend significant amounts of time immersed in the data. \reviewnew{Traditional desktop analysis systems are too complex for use in the field \cite{cutter2003gi,flentge2008designing}}. 
As a result, situated field analysis systems should balance specificity with flexibility: analysts need to rapidly generate important insights without incurring substantial cognitive overhead to build these insights. 

\noindent
\textbf{R4--Use in Outdoor Environments:} Operations in outdoor environments require analysts to carry any necessary equipment. Field analytics systems must be sufficiently portable \review{and, when possible,} robust 
to bulky equipment like gloves \reviewnew{\cite{burigat2007geographical}}.


\review{While we intended to use an iterative, user-centered design process, the lack of shared understanding about capabilities of IA made it difficult to elicit meaningful information about specific tasks and potential designs.}
To this end, we \review{used our results to build} a design probe 
to scaffold in-depth interviews about visual analytics for fieldwork.

\section{Design Probe Implementation}
\label{sec:InitialPrototype}
\begin{figure}
	\centering
	\includegraphics[width=\linewidth]{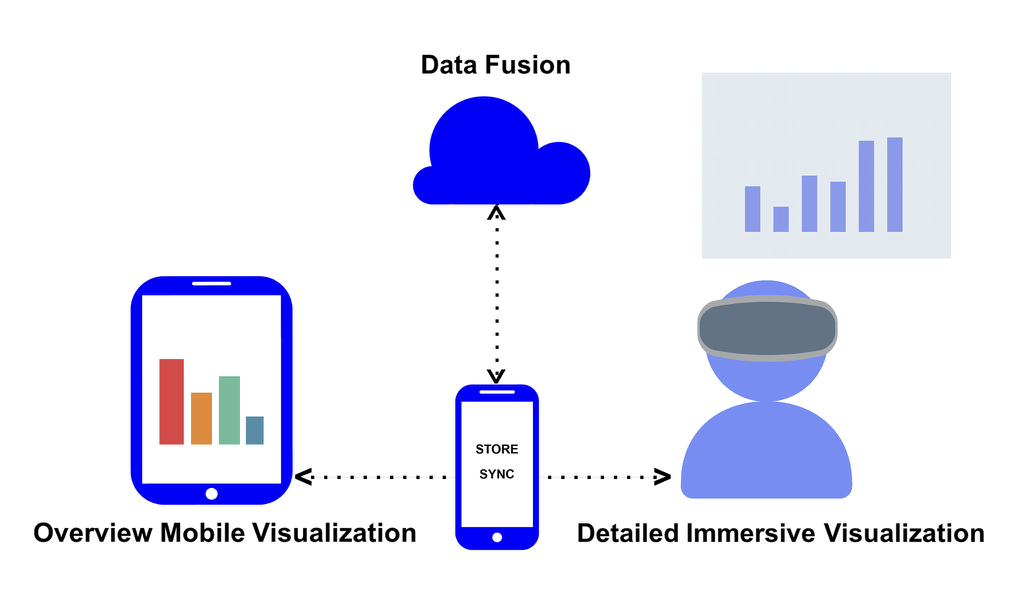}
	\caption{An overview of our 
		 design probe architecture---data is collected and merged with cached data on a mobile device and synchronized with a cloud datastore when the device is connected. Cached data can be viewed on the mobile device or explored in a portable immersive headset. 
	 }
	\label{fig:InitialSchematic_new}
	\vspace{-5px}
\end{figure}

\begin{figure}[t!]
	\centering
	\subfloat[Billboarded map in 2D on Gear VR] {\includegraphics[width=0.3\linewidth]{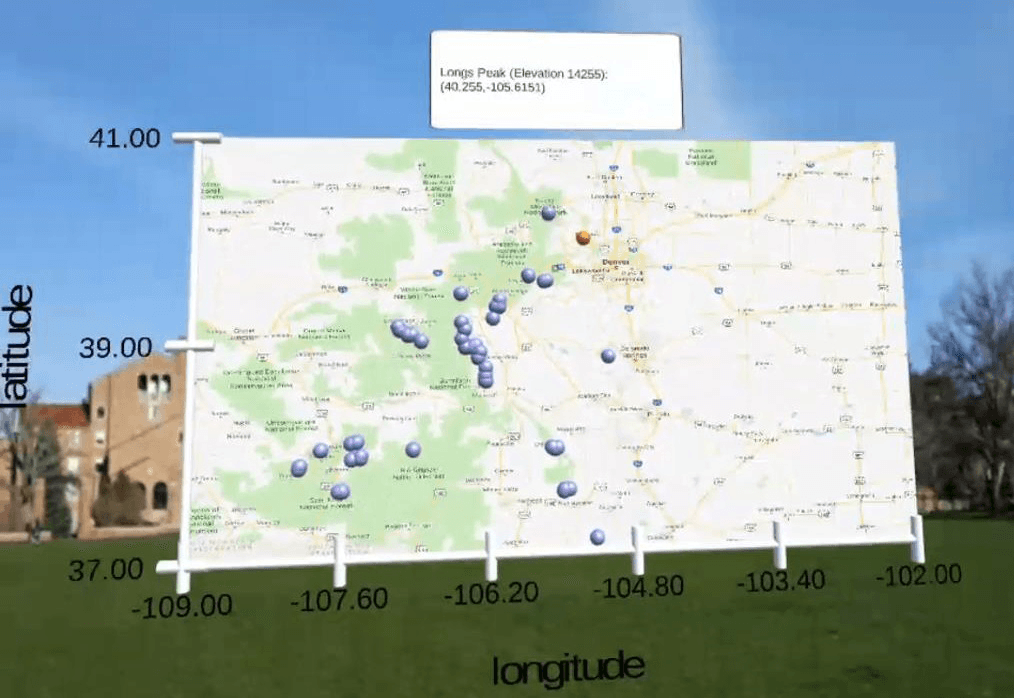}}
	\hfill
	\subfloat[3D bar chart with altitude data on Gear VR] {\includegraphics[width=0.3\linewidth]{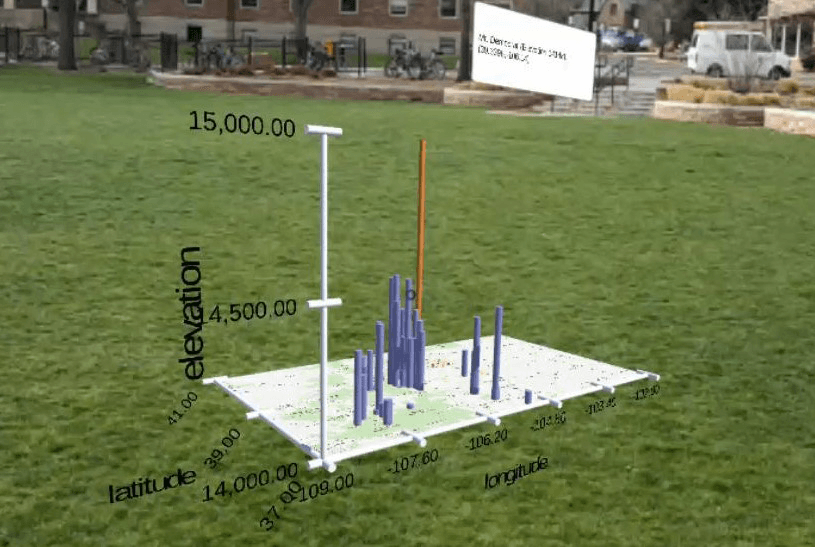}} \hfill
	\subfloat[Embedded bar chart on MS HoloLens] {\includegraphics[width=0.35\linewidth]{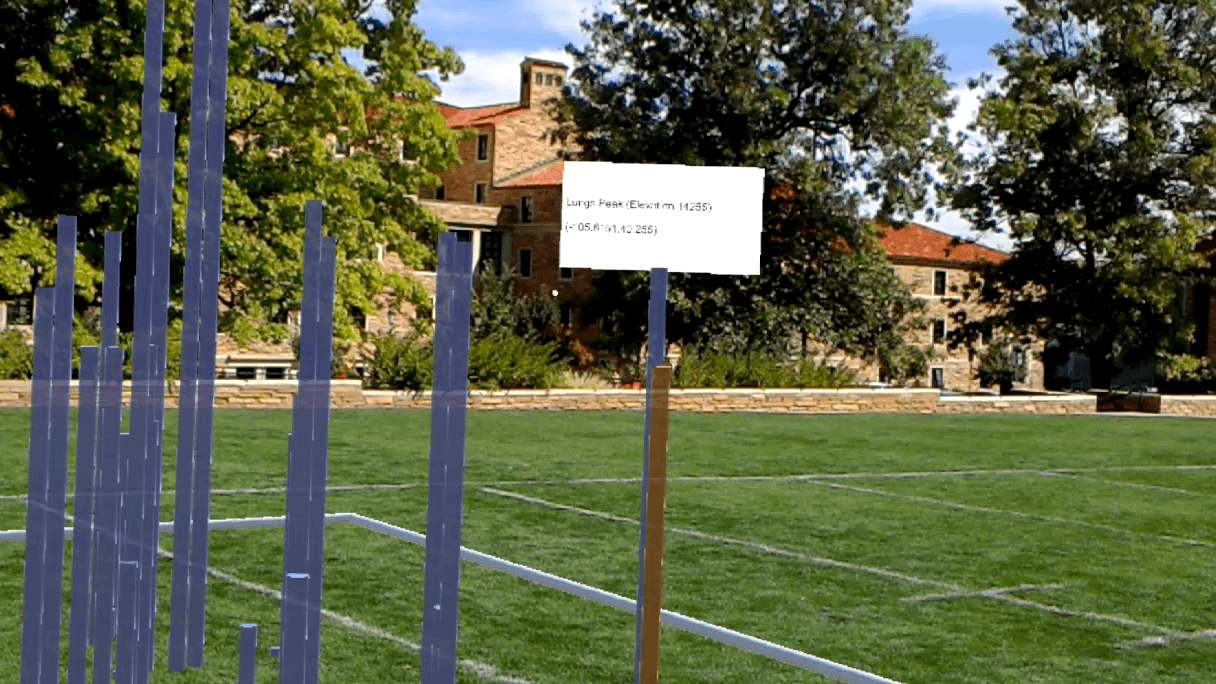}}
	\caption{Our design probe used three immersive visualizations to increase contextual awareness: \review{(a) a billboarded map, (b) a 3D bar chart, and (c) an embedded bar chart. We used simple visualizations in the probe based on R3 and to minimize bias from design in our qualitative study}.}
\label{fig:InitialImmersiveVis}
\vspace{-5px}
\end{figure}

We used \review{our preliminary interviews} to construct a design probe
to ground interviews about visualization design for fieldwork (Fig. \ref{fig:InitialSchematic_new}). 
The probe allows analysts to input data on mobile devices, synchronizing to a local database and then to a remote data store when \review{connected} (\textbf{R1}).
Analysts can visualize cached data on their mobile devices using standard overview visualizations (\textbf{R3}) or in immersive interactive visualizations (\textbf{R2}) using either a portable see-through display (e.g., Fig. \ref{fig:MissingData}) or the mobile device's camera (e.g., Fig. \ref{fig:InitialImmersiveVis}).
Our initial probe
used an Android device and Samsung Gear VR headset to \review{provide portable and integrated data collection and analysis} 
(\textbf{R4}; Fig. \ref{fig:InitialSchematic_new}). 
\review{We architected our probe using a geospatial dataset of altitude measures at various sites (seeded with the set of mountains in Colorado with elevation greater than 14,000 feet) and image data collected using the phone's camera, corresponding to data used by public safety and wildland fire experts.} 

\noindent\textbf{Target Workflow: }
We illustrate our design probe's workflow using
a scenario \review{synthesized from our preliminary interviews} where an analyst is in the field collecting geotagged data.
Analysts sync their data during preplanning, and 
can analyze the new data \review{collected during an operation} using an overview+detail paradigm distributed across devices: mobile visualizations provide rapid overviews of data collected by all teams involved in an operation while immersive detail visualizations show data related to the local environment. \review{As the analyst and other team members collect data, these visualizations update in real-time to remedy temporal analysis gaps.}
By situating data in the same physical space the data represents, 
analysts can overcome spatial analysis gaps by analyzing data in the environment. Analysts can continue this collect-and-analyze workflow over the course of the operation and refine their practices based on observations developed in real-time.

We support this workflow by integrating mobile, cloud, and immersive technologies, using an Android application for data collection and mobile visualization and immersive visualizations built in Unity.
Analysts first synchronize a remote target database with the mobile application, which extracts and displays a list of existing data \review{and creates an input form based on the preplanning schema}.
Analysts \review{navigate the target environment, adding datapoints 
to a local database that syncs to the remote database when possible.}
They can then explore this data at a glance through three mobile overview visualizations---a bar chart, scatterplot, and heatmap---to get a quick understanding of the full data corpus.
These visualizations \review{increase global situational awareness by providing an overview of data from the fieldsite.}

From within the mobile application, analysts can transition to an immersive detailed view of this data.
\review{
	The analyst sees datapoints projected into the physical environment
	using geolocation tags from the mobile device.}
These visualizations enable analysts to more fully explore the data (Fig. \ref{fig:InitialImmersiveVis}).
\review{Analysts can pull up details on demand by tapping the D-pad on the side of the headset while looking at a point of interest, and can pan the visualizations by sliding a finger across the D-pad.}
Our implementation supports the requirements enumerated in initial interviews (\S \ref{requirements}) as follows: 

\noindent
\textbf{R1--Offline \& Distributed Data Collection:} 
\review{The disparate storage media used in the field make it difficult to combine data across teams and with archival sources, requiring substantial curation prior to analysis.}  
Mobile data collection platforms can substantially improve distributed data collection \cite{tomlinson2009use,pascoe2000using}; however, limited connectivity in the field can hinder these approaches. 
We build on prior mobile data collection efforts for distributed teams and for limited connectivity.
We leverage a two-phase approach similar to Truong et al. \cite{truong2007escape} to allow analysts to locally cache new data offline and to integrate that data with active cloud datastores when connected. 

\begin{figure}
	\centering
	\subfloat[]{\centering \includegraphics[width=0.245\linewidth]{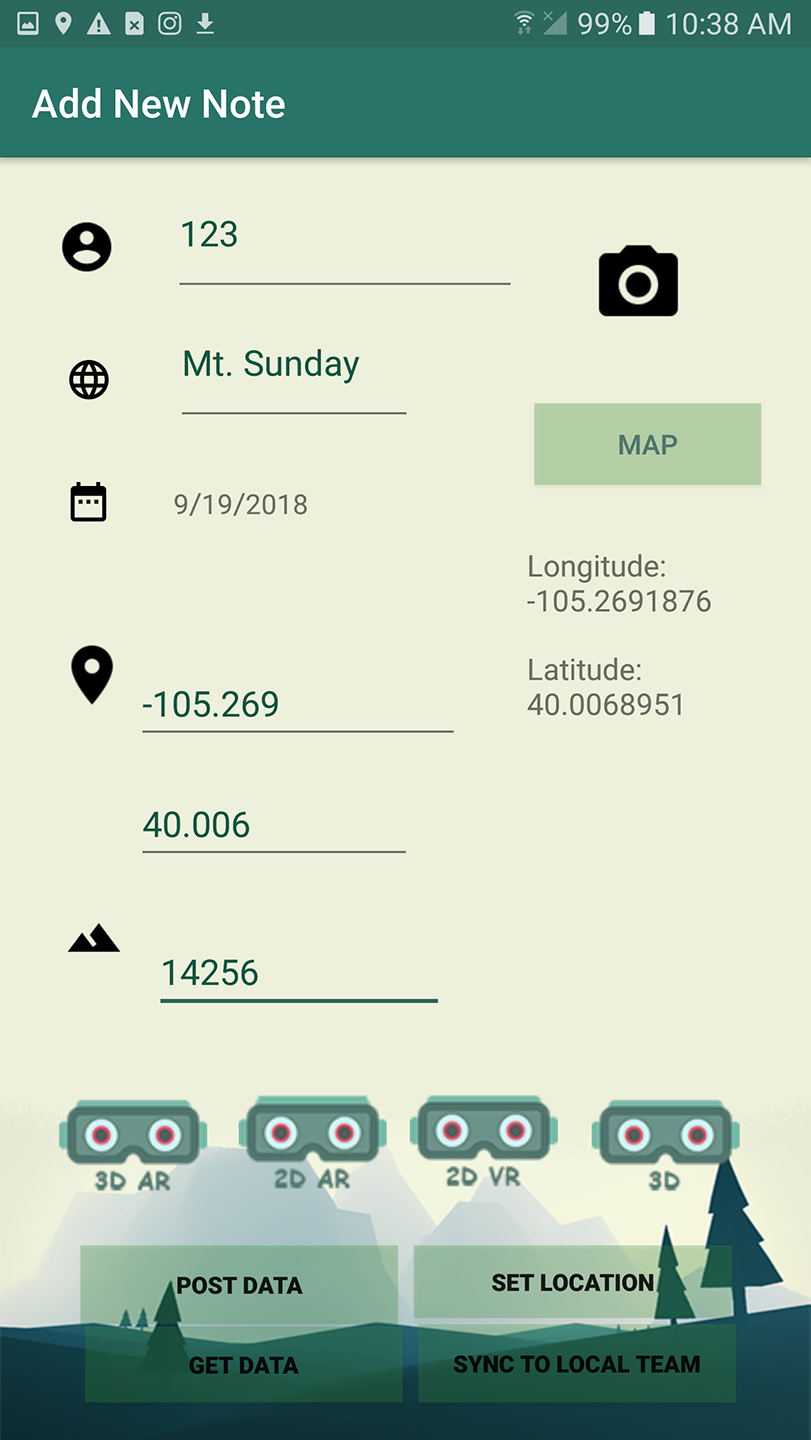} \label{fig:DataCollectionA}}
	\subfloat[]{\centering \includegraphics[width=0.245\linewidth]{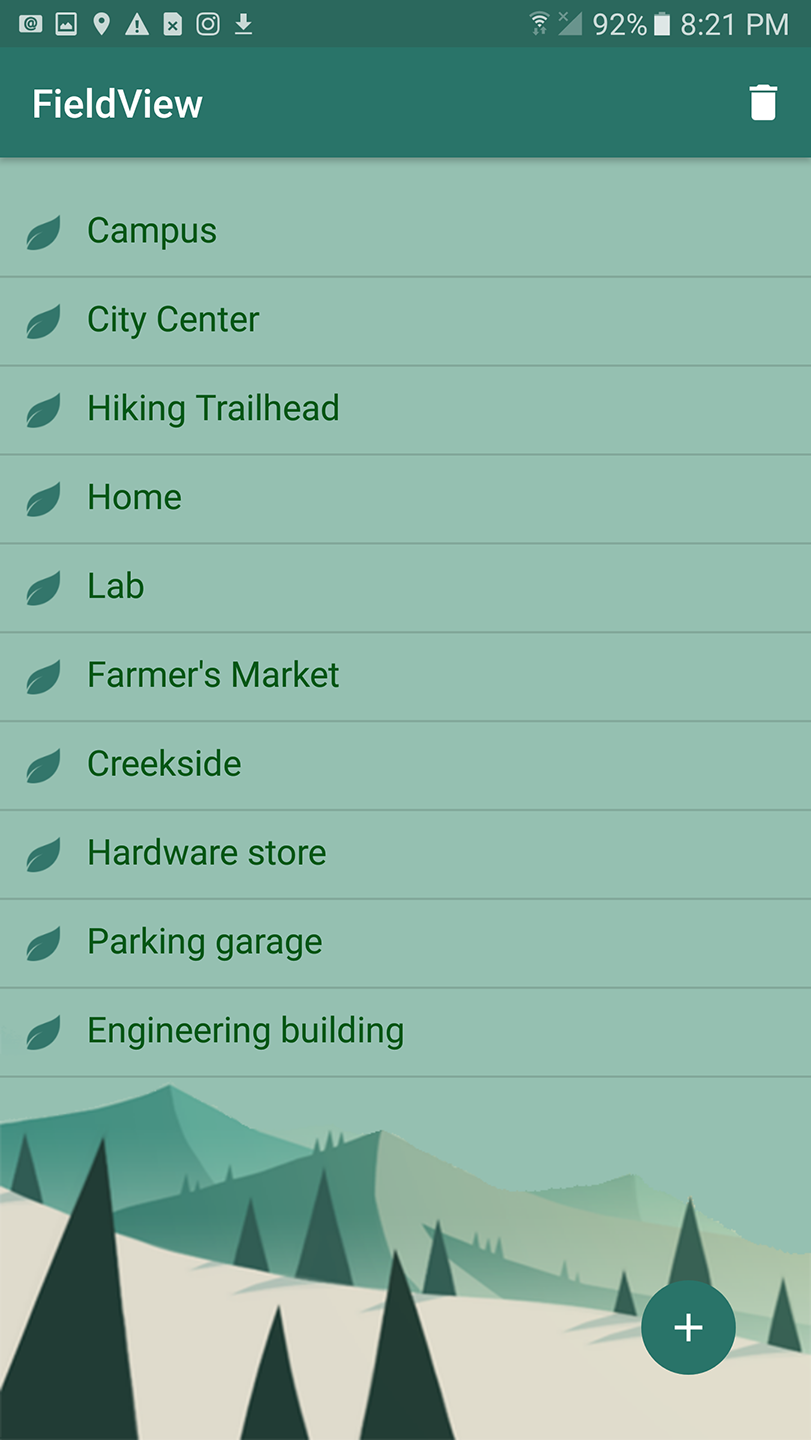} \label{fig:DataCollectionB}} 
	\subfloat[]{\centering \includegraphics[width=0.245\linewidth]{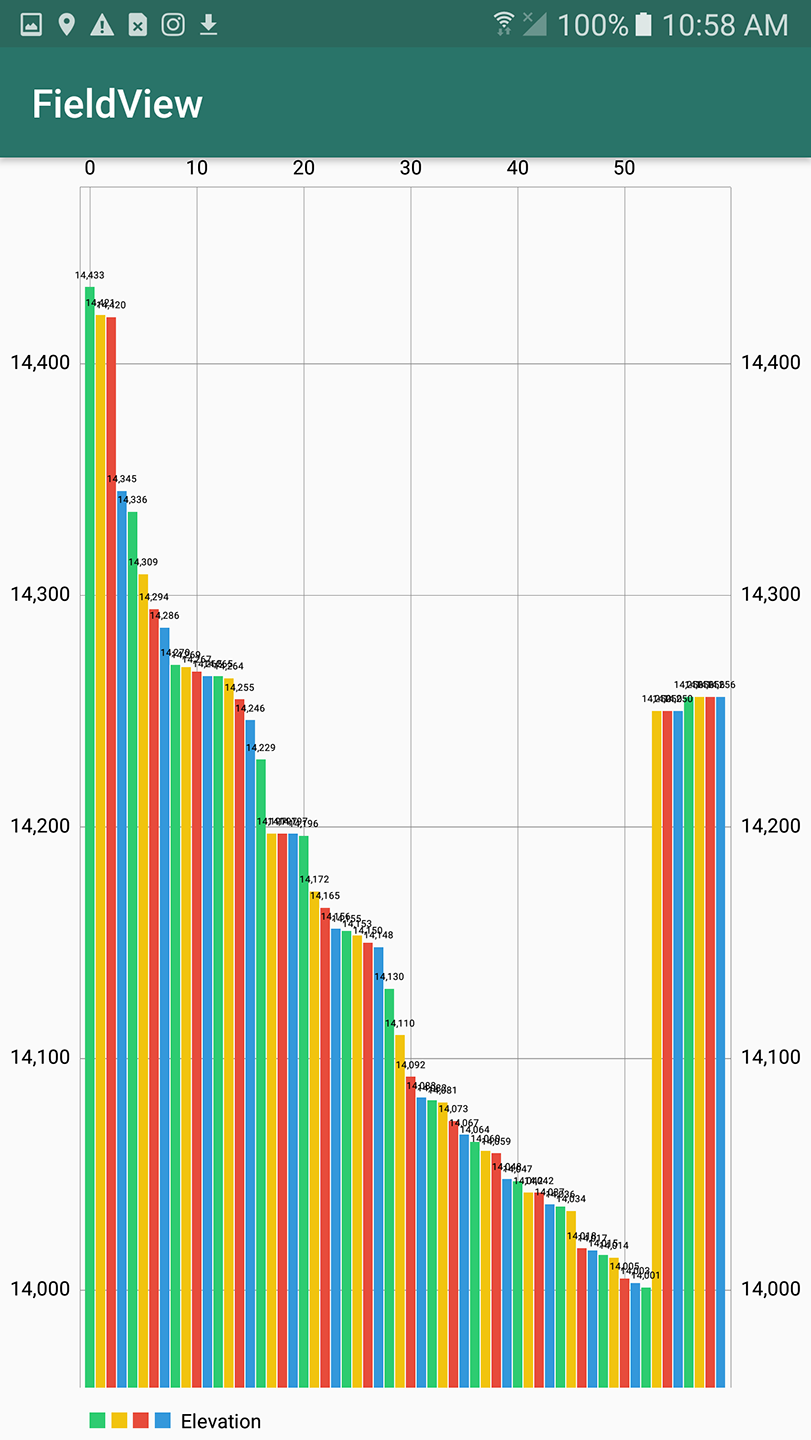} \label{fig:DataCollectionC}}
	\subfloat[]{\centering \includegraphics[width=0.245\linewidth]{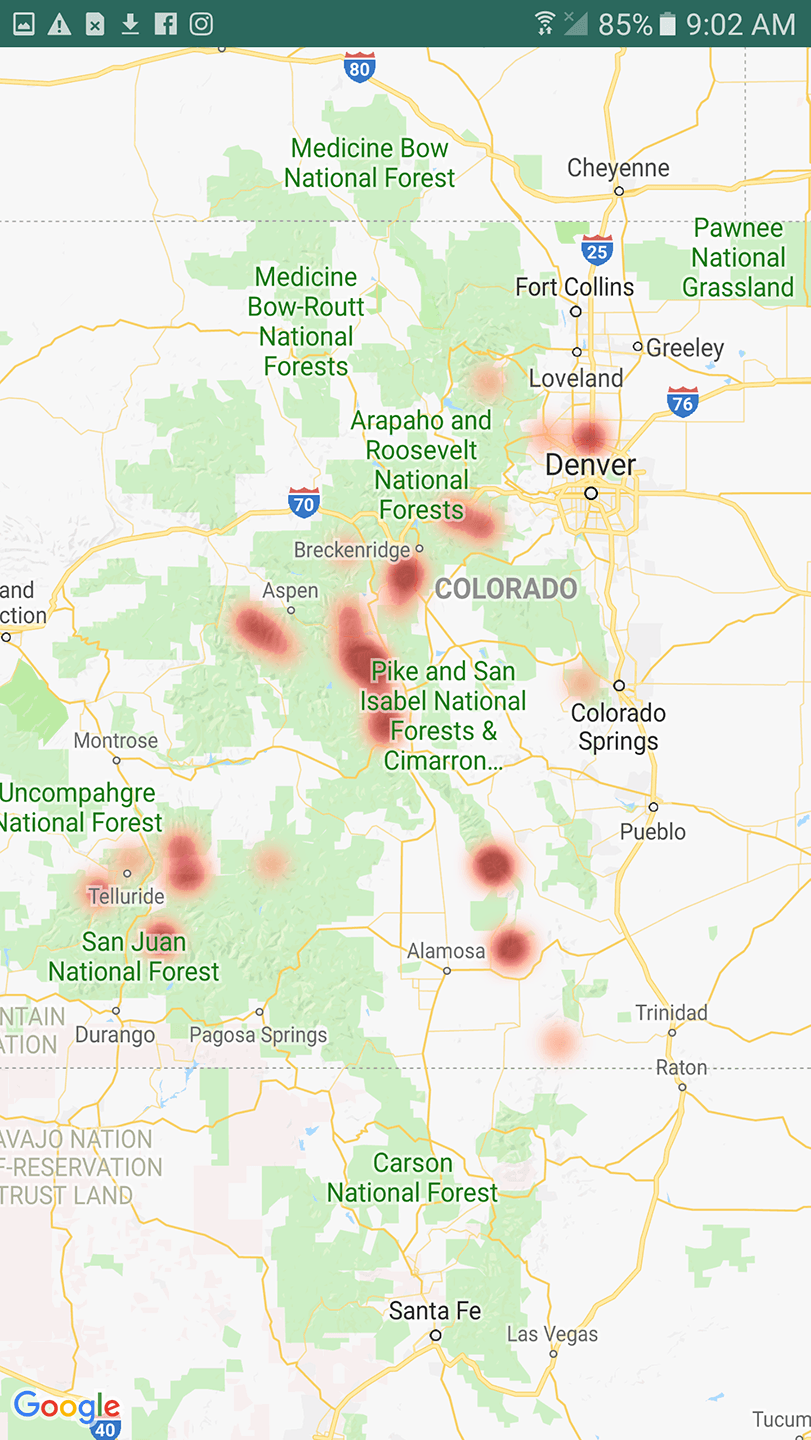} \label{fig:DataCollectionD}}
	\caption{The design probe's mobile application provides (a) a field notebook-style data collection interface, (b) a corresponding list of local data points collected in the field, and (c \& d)  interactive mobile overview visualizations based on remote server data.}
\label{fig:DataCollection}
\vspace{-5px}
\end{figure}

The application uses the remote database schema to populate fields in a field notebook interface allowing numeric, text, time, GPS, and image data collection, 
\review{standardizing} data entry across distributed teams (Fig. \ref{fig:DataCollectionA}). 
As analysts add data, new datapoints immediately sync to the local datastore. 
To support distributed collection efforts, new entries are pushed to the remote cloud database when the device is online, and any new entries in the remote database are pulled to the local datastore.
When offline, the application caches the new data and flags the data as unsynchronized to provide insight into the freshness of the available data.
To enable ready extensibility, we implement our cloud datastore using a MySQL database hosted on a standard web server. 
Our dataflow uses standard HTTP message passing where analysts could easily integrate custom server-side 
functionality tailored to particular domains or operations. 

\noindent
\textbf{R2--Merge Environmental Context \& Analysis:}
Recent advances in mobile and immersive analytics (e.g., \cite{elmqvist2013ubiquitous,willett2017embedded}) allow analysts to explore data in real time to increase situational awareness and to ground those analyses within the environmental context.
However, these techniques are currently tailored to specific applications and do not 
integrate data collection and analytics as is critical for 
field operations. 
Our probe allows analysts to launch either mobile or immersive interactive visualizations directly from the Android application. 
Mobile visualizations provide rapid overviews of the data, allowing for increased situational awareness over large geographic regions. 
Analysts can use these visualizations to identify coverage gaps in current data collection efforts. 
However, mobile visualizations suffer from 
several potential limitations 
(see \S \ref{requirements}) including decontextualization and limited screen space.

We overcome these limitations by allowing analysts to use immersive visualizations for more detailed exploration within the local environment.
These visualizations use AR to provide greater detail about data describing the immediate local environment. 
To understand how analysts might use these visualizations, our design probe uses three immersive visualization strategies:
a low-coupling billboarding approach (Fig. \ref{fig:InitialImmersiveVis}a) that provides 
a flat interactive representation of locations as scatterplot points similar to a conventional heads-up display,
a low-coupling 3D representation of the data (Fig. \ref{fig:InitialImmersiveVis}b), and
a high-coupling approach that employs spatial mapping 
to allow the user to walk around the visualizations and engage with data that keeps its physical placement as the analyst moves (Fig. \ref{fig:InitialImmersiveVis}c).
This approach of overlaying data on top of the physical environment \review{provides analysts} immediate understanding of the data's real world context by collocating data with the physical environment it describes.
\review{True high-coupling in some scenarios goes beyond spatial location to incorporate object position and structure. This mapping would require advanced computer vision techniques outside of the scope of our design probe.}

\noindent
\textbf{R3--Mitigate Information Overload:}
We couple mobile and immersive AR visualization using an overview+detail paradigm
to mitigate information overload 
\cite{ellis2007taxonomy}.
Mobile visualizations provide an overview of the data from the entire operational area while the immersive visualizations allow exploration of detailed subsets relevant to the active area of exploration.
\review{Within our mobile visualizations, analysts can quickly visualize important data dimensions using a map, bar chart, or scatterplot (Fig. \ref{fig:DataCollection}).}  
These simple designs adhere to analysts expressed needs to quickly ``sanity check'' analyses and are tailored to the limitations of mobile devices \cite{wiehr2013drivesense}, while providing simple, familiar data representations \review{to ground our discussion of the design space using this probe}. \review{To support rapid insight into local and remote data, }our visualizations encode remote data values as blue marks, cached data as green, and unsynchronized data in red (Fig. \ref{fig:InitialImmersiveVis}). 

\noindent
\textbf{R4--Use in Outdoor Environments:}
using the rear-facing camera of the phone for pass-through AR. 
We use AR rather than VR
due to the potential for AR to bind data and context \cite{willett2017embedded} and to mitigate potential occlusion issues that would inhibit active operation in an outdoor environment. 
We also implement our probe on the Microsoft HoloLens to showcase the benefits of using an AR headset with spatial mapping, notably embodied navigation and freehand gestural input.
\review{Though gestural interaction allows for hands-free data exploration, it does not resolve issues associated with data entry on a mobile device while wearing gloves.
Manual data can be input by multiple team members and the probe's architecture allows sensor data to stream to the remote database. Future iterations could explore integration of sensor-specific visualizations and verbal dictation using the phone or HMD microphone to automatically log data.}

\section{\review{Qualitative Study:} Designing for Field Analytics}


We used our design probe to ground semi-structured interviews with 10 field analysts to understand how visualization systems can bridge spatial and temporal gaps in field data practices. 
Interviews ranged from 1-2 hours and were conducted in five unique sessions.
Six interview subjects were in emergency response domains,
two were earth scientists in forest ecology, one was a computer scientist using field data for climate modeling, and one specialized in robotic sensing with drones.

\begin{figure*}
\centering
\includegraphics[width=1.95\columnwidth]{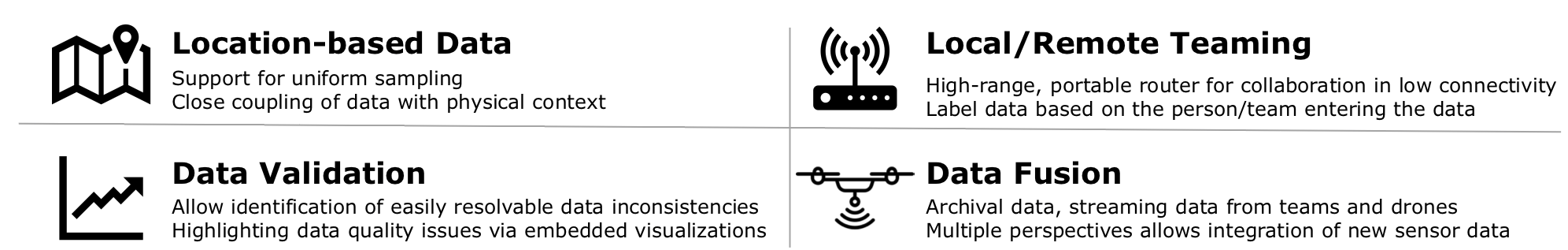}
\caption{
	We synthesized design recommendations for how data collection and visualization might enhance data use in field operations. 
	These insights fall into four primary categories that can 
	guide designers to scenarios and design \review{considerations} for field analysis tools. }
	
\label{fig:Feedback}
\vspace{-5px}
\end{figure*}

We first conducted a step-by-step interactive demonstration of the design probe's workflow with participants, starting with an overview of functionality. 
Then, we walked participants through example usage in field operations\review{---assessing terrain using the geospatial altitude dataset described in \S \ref{sec:InitialPrototype}---}interacting with the mobile application and immersive visualizations throughout this walk-through. 
We collected transcripts of comments during the participants' initial interactions with the probe and, following the demonstration, asked a series of interview questions.
Our interview questions (provided in the supplemental materials) focused on analysts' backgrounds; current use of data in the field including collection, storage, and analysis; responses to the design probe; and envisioned uses for field analysis based on the probe. 
We analyzed the transcribed interviews using thematic analysis to identify common practices in field analytics across different domains that visualization systems 
can support. 
We found that, though our participants ranged greatly in technical expertise and domain focus, they identified several common themes in potential uses for field visualization.
In this section, we discuss these themes (summarized in Fig. \ref{fig:Feedback}), how analysts envision using the design probe to support needs expressed in each theme, and how our approach could be extended to better support field analysis.
\review{Though we interviewed only 10 domain experts, we found common themes across groups that spoke to critical opportunities for visualization to enhance field practices.}
We elicit design recommendations and critical use cases for technology-mediated field analysis from these interviews.

\subsection{Reliance on Location-Based Data}
\label{locationData}
\review{All} analysts responded positively to the design probe's emphasis on geospatial data. 
Though \review{experts in each domain} leveraged different types of data, analysts generally stressed the importance of \review{spatial} data for field decision making.
For example, \review{seven} analysts used image data labeled with geospatial locations to increase global awareness of the field site.
\review{The two} earth scientists employ stratified sampling to collect data at regular intervals in a geographic area of interest
by subdividing a fieldsite into regular grids and ``taking \emph{x} number of plots in different categories'' in each gridcell to ensure sufficient sampling \emph{(P6)}.
Aerial firefighters use land ownership data defined by property boundaries to guide their operations. 
\review{The two earth scientists, computer scientist, and roboticist} leverage sensor-based or manually-entered environmental data such as temperature, humidity, glacial velocity, and tree diameter to reconstruct an environment and model spatiotemporal patterns.

The focus on maps and their fusion with the data in the immersive visualizations allows analysts to visualize data about the full space of operations. 
\review{The six participants in field-oriented domains (fire fighters, search-and-rescue, and earth scientists) remarked that the simple visualizations enabled appropriate levels of analysis while providing opportunities to dig deeper into the data through interaction. The appropriate level varied depending on the target task and context. For example, firefighters wanted access to terrain maps, ownership labels, and other archival data in preplanning, but only simple, minimally obtrustive shapes that indicated decision-relevant information (e.g., simple temperature alerts or points indicating team member or payload locations) during operations}. 
However, participants wanted to see how data could be more tightly interwoven into the physical space. 
\review{All participants emphasized that the best way to improve field practices through data would be to provide a more informed picture of the operational environment: ``the more situational awareness, the better'' \emph{(P8)}.}
This coupling could improve the integration of contextual awareness into data analysis and better support operational decision making. 

\noindent
\textbf{Design Recommendation: } Geospatial data projected onto its location in physical space could enhance data sensemaking and at-a-glance decision making. 

\noindent
\textbf{Key Tasks: } Analyze stratefied samples of local data and overhead imagery, navigate relevant property boundaries and physical structures.

\subsection{Teaming under Limited Connectivity}
\label{teaming}
All participants conduct field operations in teams. Those teams can be distributed over a wide space (e.g., around the perimeter of a wildfire) or collocated (e.g., collecting tree cover measurements in a specific fieldsite).
Currently, \review{all noted} data sharing between teams is done through direct communication in collocated teams or by radio commands from an operations center.
None 
had solutions for sharing and collaboratively analyzing updated field data across teams in real time.

\review{All but the roboticist (9/10 partipants)} were excited about the potential of analytics tools to support teaming in both collocated and distributed environments and support consistency checks across team members. 
For example, the modeling expert noted an incident where two teams at a fieldsite calibrated their sensing equipment differently, leaving their data unusable. 
Shared data between these team members would allow for at-a-glance identification of these potential anomalies. 
For distributed teams, aerial firefighters noted that recent advances have allowed them to share basic geocoordinates alongside radio transmissions to help improve situational awareness, but that these techniques are extremely limited. 
They anticipated labeling data values based on team entry in the map visualizations would significantly expand their awareness of the state and progression of a fire and let them more readily adapt operations through communication between teams.

\review{Participants in emergency response and climate modeling} saw the probe's two-phase caching scheme as beneficial in some instances, but insufficient for scenarios 
where the entire fieldsite lacks connectivity.
In these sites, even collocated team members may struggle to share data.
Our search-and-rescue expert observed that ``[rescue workers] all have cell phones and we can get so much data until we don't have cell service and then you can't do anything'' (\emph{P8}). 
The climate modeler typically uploads data and receives an automated email with summary statistics to confirm the data has been stored. 
\review{When mobile on a fieldsite, this confirmation pipeline fails making} 
	``a validation of field data...the first thing to try to support'' \emph{(P9)}. 
Aerial firefighters noted a willingness to carry a high-range portable router to collate data from collocated teams, similar to the peer-to-peer infrastructure in Siren \cite{jiang2004siren}. Field analysis systems may be able to use \review{this approach for}
visualization to improve data sharing among local teams.

\noindent
\textbf{Design Recommendation:} Integrating intermediate shared data caching can support collocated teaming in environments with no connectivity through the visualization of both cached and cloud data.

\noindent
\textbf{Key Tasks: }Calibrate between sensors and teams, increase communication and global awareness between teams.

\subsection{Data Quality Validation}
\label{quality}
Participants 
advised against attempting all operation center functionality in AR.
For example, one earth scientist stated ``when I am engaging in the data in a complex way... I am at home at my computer. The things I want to know in the field are pretty simple'' \emph{(P6)}.
Interviewees emphasized the need to avoid overwhelming analysts with too much data while in the field \review{and to focus on operational tasks that analysts can assess and address in context such as validating the quality and consistency of incoming data.}

\review{Nine analysts} noted that data collection errors force them to return to the field and repeat data collection if errors are found while at the fieldsite or to discard large amounts of data if errors are undetected until later analysis.
For example, our modeling expert remarked that mistakes in data collection for glacial research were unresolvable as they required flying equipment back to remote locations in Greenland or Antarctica.
Earth scientists remarked that measurements of tree circumference---typically taken at chest height---could introduce substantial variation depending on the scientist's height. 
While these inconsistencies are easy to resolve, they are currently impossible to detect in the field. \review{Four analysts specifically mentioned} validating data by ``looking at the size of the file and saying ``that looks about right'' \emph{(P10)}.


Analysts saw the design probe as a way to immediately evaluate collected data for errors in quality, \review{with the four participants in the sciences} noting that these visualizations would save substantial time and money in field operations. 
They further remarked that the simplicity of the probe's visualizations would allow analysts to quickly locate errors using familiar representations and to verify consistency by exploring high-level patterns in existing data. 
Quick glances at the mobile visualizations provide a sense of data coverage, while the immersive visualizations allow analysts to dig deeper into anomalies and compare those variations side-by-side with the conditions of the environment to remedy collection issues. 

\review{Four emergency responders and the climate modeler} noted that systems may further aid in data quality validation by automatically highlighting potential quality issues, a challenge that could be addressed by coupling automated and visual analysis. 
Analysts recommended field analytics systems more deeply consider how embedded visualizations might aid in quickly identifying problematic areas in data collection to increase data quality and efficiency. 
Such technologies could allow analysts to actively update their goals and operational strategies as a function of the quality and completeness of incoming data. 
 

\noindent
\textbf{Design Recommendation: } Analysis systems should explicitly support detecting, contextualizing, and remedying low quality data.

\noindent
\textbf{Key Tasks: } Identify missing data, detect anomalies, monitor incoming data.

\subsection{Data Fusion Across Perspectives}
\label{fusion}
\review{All ten field analysts} rely on archival data, such as environmental surveys, topological maps, or data from prior efforts, to plan each day's operations. 
They noted that the design probe's visualization systems could improve efficiency by bringing these sources into the field.  
For example, aerial firefighters wanted to overlay land ownership markers on the operational site.
Public safety officials wanted to see key geological features, such as altitude shifts. 
Earth scientists wanted to couple visualizations of tree health from prior field surveys to study changes over time.
Updating visualizations during data collection may allow analysts to inform operations and change data collection practices on the fly. 
\review{Nine of the ten participants} felt that extending the immersive visualization suite to integrate other teams' locations could further improve operational efficiency, allowing analysts to better coordinate operations across teams collecting data over a single site. 

While the design probe relies heavily on archival and manually collected data, analysts wanted to also fuse this data with data collected by robotic drones and other physical sensors.
\review{All participants saw the distributed notebook interface as an improved method for merging data across teams in many scenarios; however, the emergency responders and earth scientists felt the notebook collection interface was potentially problematic when gloves were required for safety or due to environmental conditions (e.g., working in active burn zones or tropical bogs). Integration with automated sensors helps overcome these limitations and expands both the data variety and field coverage used in analysis: sensors can stream more data faster and from more locations.}
In wildland firefighting,
sensor readings of temperature, wind, and humidity significantly aid analysts in estimating wildfire spread, but ``the more data [analysts] get, the harder it is to process and analyze and make decisions off of'' \emph{(P2)}. 
Experts noted that the design probe's approach could significantly improve the integration of such sensing devices by allowing analysts to fuse archival data with incoming data streams currently only accessible to operations centers. 
For example, 
image data collected by drones can provide multiple perspectives
that extend the area an analyst can see. 
In earth science, this integration would allow analysts to simultaneously study canopies from above and below to estimate forest health.
All analysts noted how practices could be improved by visualizing sensor data in real-time, 
and that visualizations of this data would bring new kinds of data into the field.
Our search and rescue expert said, ``there's only so much you can learn from a picture or video... When someone provides me with a cool sensor and I can get real-time data on gas concentrations on a hazmat incident and I can have it broadcast in, I'll be stoked'' \emph{(P8)}.
Analysts recommended that fused views should also allow for separating data from different sources, giving field analysts the ability to distinguish preplanning data from data collected by team members and autonomous sensing technologies.

\noindent
\textbf{Design Recommendation: } 
Visualizations should incorporate data from other sources, including autonomous drones, to increase awareness of both local and global operations.

\noindent
\textbf{Key Tasks: } Couple data and images from multiple sources, stream real-time data from autonomous sensors into the field. 

\section{FieldView} \label{sec:design_iteration}
\begin{figure}[t!]
	\centering
	\includegraphics[width=\linewidth]{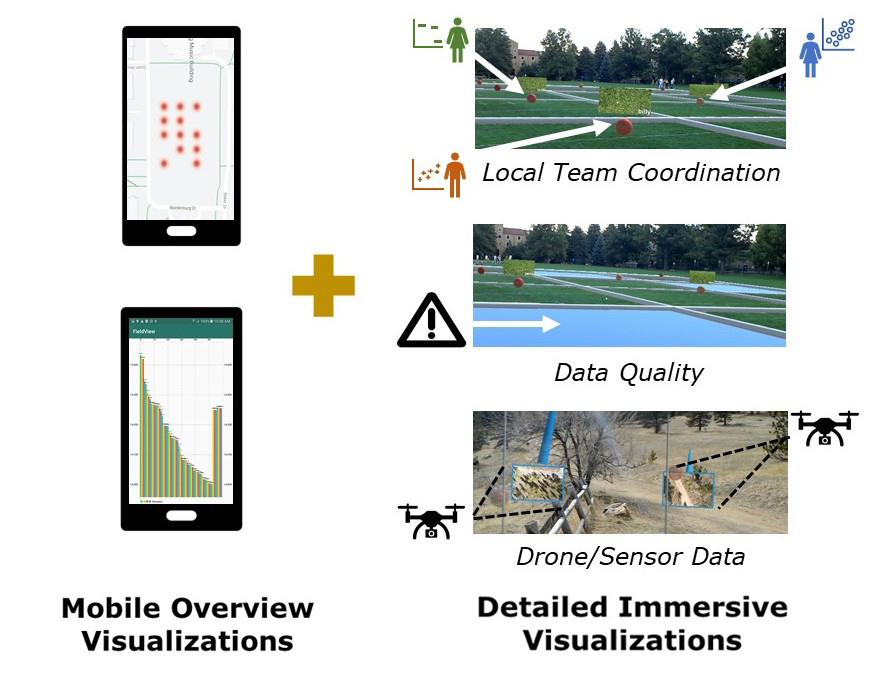}
	\caption{We \review{iterated on our design probe using expert feedback to create the FieldView prototype}.
		FieldView shows how visualization can support three use cases from our interviews: localized team coordination, data quality validation, and data fusion from heterogeneous sources.}
	\label{fig:UseCases}
	\vspace{-5px}
\end{figure}
\review{Our interviews revealed design recommendations and key scenarios for how visualization can enhance field practices.} 
To provide a proof-of-concept for how visualization systems might \review{use these recommendations},
we developed FieldView, an open-source extensible prototype, on top of our design probe's base implementation. 
\review{FieldView extends the probe's data collection functionality through an intermediate cache on a portable server to support collocated teaming under limited connectivity (see \S \ref{teaming}) and basic sensor data integration (see \S \ref{fusion}). It includes situated geospatial AR visualizations designed for three operational tasks noted in our interviews: 
collocated teaming, data quality validation and refinement, and autonomous sensor integration.}
As there is little empirical guidance for immersive visualization design (see Marriot et al. for a survey \cite{marriott2018immersiveCH2}), 
these use cases demonstrate how situated analysis may enhance field operations and decision making and illustrate novel capabilities of our system 
to support field analysis (Fig. \ref{fig:UseCases}). 

Case 1 and Case 2 simulate multiple analysts collecting scorch rate data, drawing on a use case from 
our interviews with ecologists (``sampling what percentage of needles are scorched''\emph{(P6)}). As we were unable to revisit the original fieldsite, we retargeted the geocoordinates of the scorch rate data to a local fieldsite.
Case 3 simulates a desired ``synthetic vision'' (\textit{P1}) provided by seeing the physical environment with temperature data and drone-collected imagery \review{provided by our earth science collaborators and again retargeted to a new fieldsite}. 
We published these use cases as part of an open-source system designed for use with these cases and readily extensible to new data and devices based on the users' operational needs (\url{https://cmci.colorado.edu/visualab/fieldview}). 

The Gear VR lacked robust tracking,
limiting our ability to integrate data into the environment. While these capabilities can be gained through additional sensing devices, many recent headsets have these capabilities readily available. 
Our case studies use the Microsoft Hololens with freehand gestures for 
\review{interaction} as \emph{P1} reiterated challenges with gloves while in the field inhibiting touch interaction.
\review{However, as with the design probe, FieldView does not address the challenges of entering data on a mobile device while wearing gloves.}
Our modular architecture (Fig. \ref{fig:InitialSchematic_new}) allows analysts to extend FieldView to specialized use cases and technologies---such as other AR headsets or more sophisticated fusion and analytics algorithms---by changing the data source and schema and integrating custom analysis processes and visualizations.

\subsection{Case 1: Team Coordination}
\begin{figure}[t!]
	\centering
	\includegraphics[width=\linewidth]{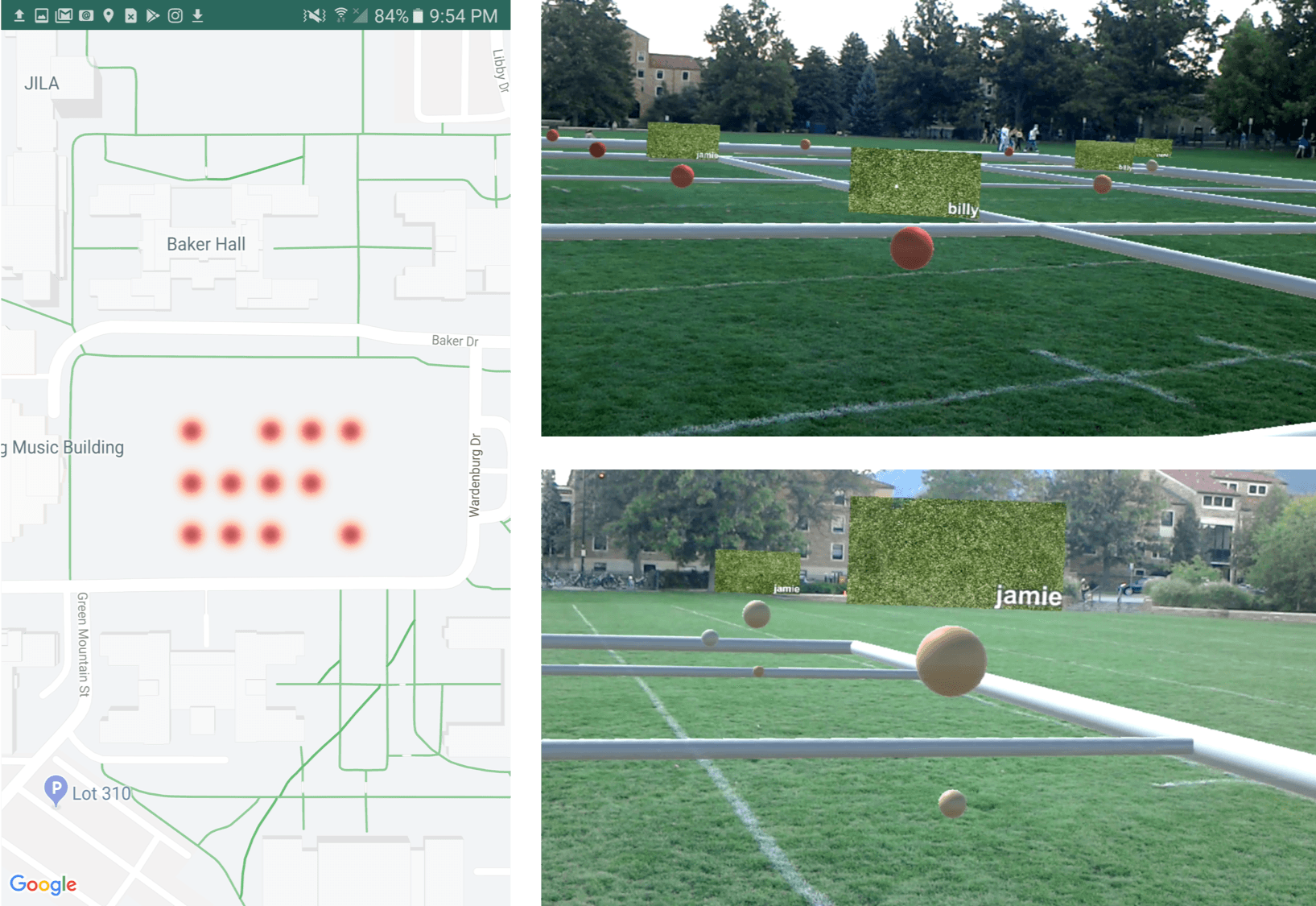}
	\caption{Use Case 1: Team coordinators can assess coarse-grain coverage at a glance on the mobile device (left). They can survey team responsibilities and validate collected data (redder data points indicate greater scorch percentage) using a contextualized scatterplot (top right) and \review{compare new data to archival data (bottom right).}}

	\label{fig:LocalTeamCoordination}
	\vspace{-5px}
\end{figure}

Interviewees discussed the potential for the data collection and immersive visualization components of the design probe to support team coordination
by aggregating data collected by multiple team members (\S\ref{teaming}) and enabling a shared situational awareness across sites.
While remote fieldsites may complicate use of mobile devices and cloud infrastructures, our public safety and aerial firefighting experts were willing to carry lightweight 
portable servers that could establish a local connection between collocated teams and mentioned using ATAK\footnote{http://syzygyintegration.com/atak-android-tactical-assault-kit/} servers for basic data exchange. 
\review{In this use case, field analysts can enter and synchronize data with local teams while offline (Fig. \ref{fig:LocalTeamCoordination}).}

A local Flask server is hosted on a Raspberry Pi (23 grams) and connects analysts to a local network.
Team members use FieldView's data collection interface to add new scorch rate samples alongside image references drawn from archival data. 
\review{All analysts connected to the server can add new data to a local database that synchronizes with the server's intermediate database while within range, allowing the entire team to view data collected within range of the server even when no internet connection is available} \review{(\S \ref{teaming})}.
\review{When a team member enters a new scorch measure, the visualization updates and all team members connected to the server have access to the new data.}
Team members engaged with the mobile platform can quickly view coverage of a surveyed area on the mobile application's map view, 
while analysts engaged in immersive analysis can see incoming data overlaid on the field (Fig. \ref{fig:LocalTeamCoordination}). 
Our overlay visualization uses GPS data from the analysts' mobile device and the HoloLens' internal depth sensors to localize data over an explicit collection grid, supporting stratified sampling \review{(\S \ref{locationData}).}
Outdoor and large-scale tracking issues with current AR headsets and mobile GPS granularity can limit the utility of this approach, but analysts mentioned they generally use more precise GPS devices that help mitigate these limitations. 

\review{Analysts first explore the 
	mobile visualization to get an overview of data coverage with a simple heatmap of the sampled area.
They can move to a relevant field location and launch the immersive visualization to see where data was collected within that site and to map teams in the field to operational goals.}
Within the immersive view, analysts can visualize scorch rates using an embedded scatterplot applied to the field grid with points color mapped to scorch intensity values (Fig. \ref{fig:LocalTeamCoordination}). This approach minimizes the amount of space consumed by the visualization while contextualizing data in the environment and supporting stratified sampling practices.
Analysts can select collected data points \review{via freehand gestures} to display archival image data collected at that location and the \review{team member} who uploaded the data on demand. \review{As new data is uploaded, older data from the same grid location forms a smaller point underneath the most recent measure to track changes over time.}

This prototype extends current field analysis capabilities to address three design considerations identified in our interviews.
First, the dynamically-updated gridded scatterplot paired with  archival data increases the local team coordinator's awareness of the team's operations across the field site
(\S\ref{locationData}).
With an updated data model and contextually relevant data situated in the physical environment, field analysts can visualize coverage and synchronize efforts across limited human resources in real time. 
Second, 
new data is combined with preplanning data in real time, allowing analysts to update and extend data in existing databases based on mobile inputs (\S\ref{fusion}).
\review{Finally, the simple, embedded scatterplot provides at-a-glance insight into collected data to allow the team to adjust and adapt their operations in response to incoming data. }

\subsection{Case 2: Data Quality Validation}
\begin{figure}
\centering
	\includegraphics[width=\linewidth]{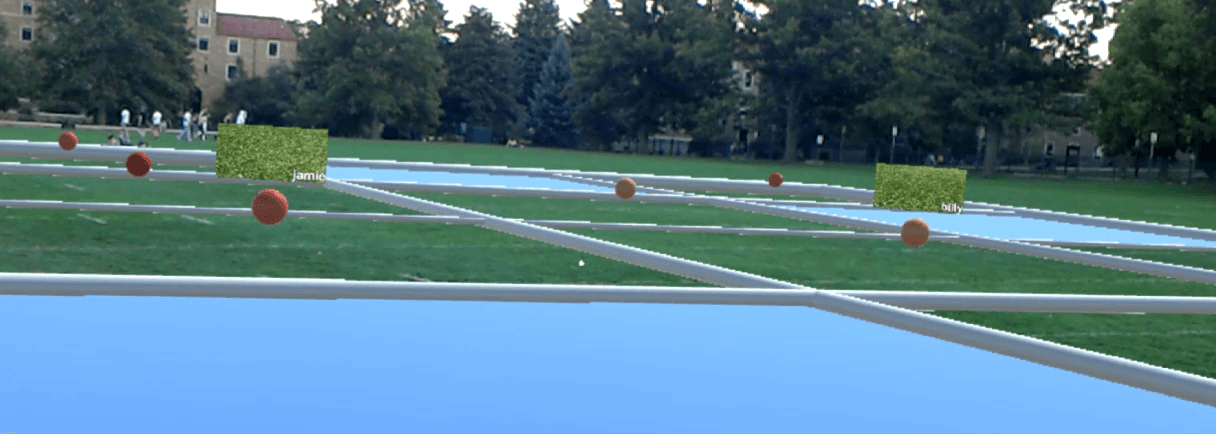}
\caption{Use Case 2: 
	Scorch data is visualized in a scatterplot embedded within the field sampling grid with one point per grid to encode scorch rates, mitigating occlusion of the surrounding environment. Blue regions within the grid allow analysts to rapidly identify gaps in data coverage.}
\label{fig:MissingData}
\vspace{-5px}
\end{figure}

Incomplete and poorly collected field data costs analysts time and money, but 
is difficult to detect while in the field despite analysts being well-positioned to correct the errors. 
Automated methods for detecting these errors are limited, and current mobile visualizations do not provide sufficient resolution to effectively detect most errors in the field. 
Further, understanding the context of collected data can enable rapid identification and correction of data collection errors. 
For example, detecting anomalies while in the field can call analysts' attention to sensors that may be occluded due to weather or dirt or that may be improperly configured. 
By visualizing data in the context of the operational area, field analysts can identify and rectify these errors as they occur, leading to
substantial savings in both time and budget.    

\review{Analysts can validate field data in FieldView at a global scale using a mobile map overview visualization (Fig. \ref{fig:LocalTeamCoordination}) to identify regions with missing or unusual measures. They can then move to that region and} 
locate 
missing data highlighted on a AR detail visualization (Fig. \ref{fig:MissingData}).
FieldView highlights gridcells without collected data in blue \cite{songs}, while existing data are encoded in the gridded scatterplot from \S6.1.
\review{When the analyst enters data for the missing region in the mobile app, they see the blue highlight replaced with an appropriately color-coded datapoint in real-time.}
When coupled with the offline teaming infrastructure from \S6.1,  this functionality allows analysts to coordinate data validation efforts with team members performing distributed data collection.

\review{Analysts can also use FieldView to explore patterns across an environment to look for anomalies that might suggest stale data.}
For example, analysts can survey scorch rate data across the fieldsite and compare the distribution of sphere colors against the 
expected data distribution of the visible environment to find inconsistencies.
Once those inconsistencies are located, analysts can walk directly to any problematic areas, assess the source of the inconsistency (e.g., by using imagery data from Case 1), and update data. 


This use case addresses \review{data quality assessment tasks mentioned by all participants:}
data completeness and correctness (\S\ref{quality}).
Visualizing data within the environment allows analysts to readily bring expertise and contextual information to bear on on-going data collection to identify and remedy simple anomalies. Future extensions of these approaches could integrate automated or comparative solutions into quality analysis. For example, our modeling expert envisioned integrating these approaches with automated processing algorithms used by glacial researchers to call attention to salient irregularities in their data. Forest ecologists envisioned direct visualizations of statistical power metrics overlaid on the physical environment to assess anomalies based on data density and other measures. 
Our approach enables analysts to rectify errors by providing the means to identify the presence of anomalies in mobile overviews and detailed investigation using IA.

\subsection{Case 3: Data Fusion Across Perspectives}

\begin{figure}[t!]
	\centering
	\subfloat[Overhead view of distant imagery with a wedge object as for geospatial reference]{\centering \includegraphics[width=0.47\linewidth]{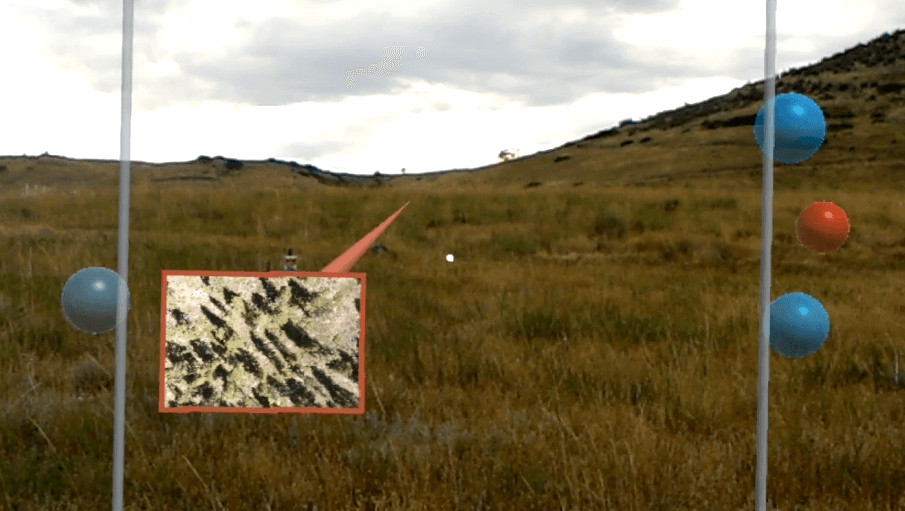} \label{fig:DataFusionDistant}}
	\,\,\,\,
	\subfloat[Overhead view of close range imagery with a dotted dropdown object for geospatial reference]{\centering \includegraphics[width=0.47\linewidth]{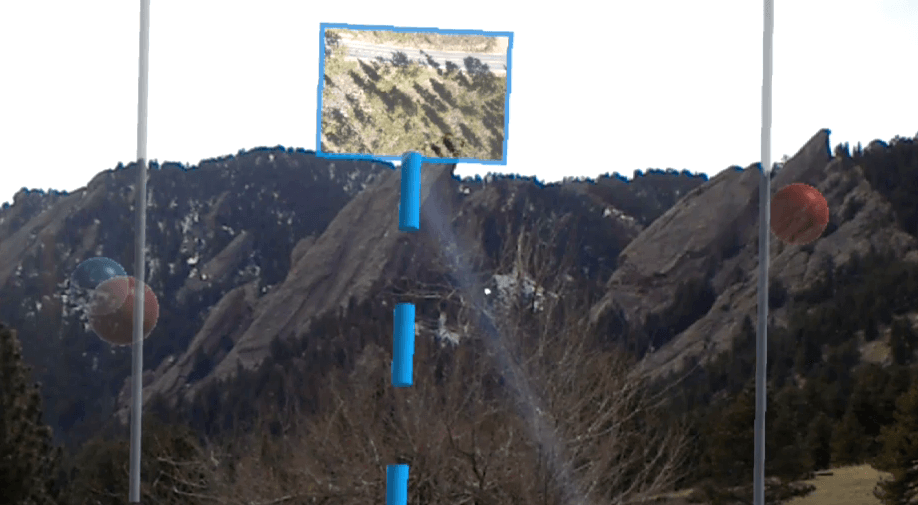} \label{fig:DataFusionNear}}
	\caption{Use Case 3: Visualizing geotagged sensor data (temperature encoded by color) and imagery
		allows 
		analysts to leverage autonomous sensing data in real time.
		We complement situated visualizations with a heads-up display (HUD) in the left and right peripheries to help find datapoints beyond the line of sight.}
	\label{fig:DataFusionUseCase}
	\vspace{-5px}
\end{figure}

Autonomous data collection is increasingly central to field data \review{and can help mitigate data entry challenges due to bulky equipment such as gloves}.
Aerial firefighters want to extend their own vision with imagery captured by drones and supplement that imagery with additional measures collected using on-board sensors; earth scientists wish to examine a forest canopy from both the air and the ground
 (\S\ref{fusion}).
Analysts can use FieldView to explore autonomous data both within the field of view and from nearby devices. We demonstrate these visualizations using a dataset of overhead imagery from drones used by our earth science collaborators during a recent field operation coupled with synthetic temperature data. 
Field analysts can use these visualizations to increase their situational awareness using data from autonomous devices and to coordinate action between human and robot teams.

\review{Analysts can visualize distant imagery through an adapted Wedge technique \cite{gustafson2008wedge}.
Within the immersive visualization, analysts see drone data \review{collected within their field of view} as scatterplot points positioned at the points of collection, with color of the image border and wedge object encoding temperature using a red-blue color scale interpolated in CIELAB.
They can then select a point via freehand gesture to see the source image, which connects the image to its physical referent using a color-coded wedge
\review{\footnote{Stereo viewing in AR-HMDs allow these points to appear at the point of collection, but positional depth cues are lost when rendered to a 2D figure.
	} } (Fig. \ref{fig:DataFusionDistant})}.
\review{To avoid information overload and keep the center of the analyst's view unoccluded, images render on either the left or right side of the field of view.}

\review{To provide more comprehensive awareness of local conditions, analysts can also view data from devices outside their immediate field of view}
\review{
	via a heads-up display (HUD) along left and right periphery.}
To keep these visualizations simple (\S\ref{quality}) and to further minimize any unnecessary occlusion (\S \ref{requirements}),
data in the HUD is rendered using a variant of SidebARs \cite{siu2013sidebars,bork2018towards}: data from each collection site is visualized as a semi-transparent point, with points projected along a vector from the measurement point to the viewer. 
The color of each point corresponds to the current temperature reading. 
This design mirrors the needs of wildland firefighters who note that the intensity and spread of a wildland fire can change quickly and having data from remote sensors will enable them to react to changes in the environment.

Our interviews found that data fusion with both human-collected and autonomous sensors providing
multiple perspectives on a fieldsite is 
key for increasing situational awareness beyond current capabilities.
Our design choices aim to balance simplicity and situational awareness for field analysts engaging with automated sensors. 
While our approach builds on techniques for visualizing data beyond the immediate field of view, future techniques could more deeply consider how visualizations could enable collaboration between analysts and autonomous systems. 
\section{Discussion \& Future Work}

This work 
\review{provides formative insight into how visualization systems can enhance field work,}
focusing on solutions that integrate data collection and analysis through combined mobile and immersive visualizations.
We developed 
a design probe to ground 
interviews with ten domain experts and identified four key design themes and several analysis tasks relevant to field visualizations.
We embodied these themes in our FieldView prototype
to provide a proof-of-concept of how visualization might enhance field work in three scenarios: \review{teaming, data quality validation, and visualizing autonomous sensor data.}
Our results provide a formative basis for designing visualization tools for field analytics and raise new questions for future tools, including: 
\vspace{-6px}
\begin{itemize}
	\item What kinds of analysis and decision making processes should field analytics tools support? \vspace{-6px}
	\item How can we optimize visualization designs distributed across mobile and immersive technologies? \vspace{-6px}
	\item How can we design integrated solutions for effective collection and analysis workflows in the field? \vspace{-6px}
	\item \reviewnew{How can visualizations better consider hardware constraints?}
\end{itemize}
\vspace{-6px}

\textit{What kinds of analysis and decision making processes should field analytics tools support?} 
Our interviews revealed common tasks that field analysis tools can support, such as evaluating data quality, supporting distributed teams, and integrating data from drones and other automated sensors. 
These interviews also yielded critical design recommendations for \review{visualizing this} data.
For example, visualizations should incorporate archival data with collected data, 
display this data 
in its physical context, mitigate information overload, and consider constraints of outdoor environments.
FieldView provides a proof-of-concept for how systems might reconcile the spatial gap by visualizing diverse data in the physical environment it was collected.
This approach also helps alleviate temporal gaps by allowing analysts to make timely decisions while out in the field.
\review{Our interviews uncovered} scenarios, datasets, and components of visualization designs that can support field analysis, but we still lack empirical insight into which tasks require field support and which tasks are best left to the command center. 
Future investigations into specific domains may provide frameworks for 
\review{optimizing} such distributed sensemaking.

\textit{How can we optimize visualization designs distributed across mobile and immersive technologies?}
Evolving AR, robotic, and collaborative technologies offer new challenges for effective data visualization, especially in scenarios where 
data comes from heterogeneous sources.
Immersive AR allows analysts to utilize their implicit sense of the environment (e.g., the trajectory of wildfire smoke or the physical bridge being inspected) alongside data.
Mobile solutions enable field analysts to engage with complex visualizations in the field, but all experts emphasized the need for understanding situational data at a glance.
Analysts needed both quick overviews of the data and the ability to make sense of geospatial data in more detail supported by physical context.
Positive feedback from the design probe 
suggests significant potential for integrated 
mobile and immersive visualizations; however we have little understanding of how to optimize visualizations to leverage these differing form factors simultaneously.
Future research should further explore 
combined mobile overview and detailed immersive visualizations, similar to the David \& Goliath approach to complimentary smartwatch and large display visualizations \cite{horak2018david}.

\textit{How can we design integrated solutions for effective collection and analysis workflows in the field?}
With our design probe and FieldView, we built example workflows that could 
empower field analysts with visualizations for data-oriented decision making.
Despite significant excitement about how field visualization tools could transform existing practices, we were limited in our ability to evaluate FieldView in the wild due to scientists' reluctance to take new technology into the field, stemming from the high time and financial costs of field operations and legal restrictions around head-mounted displays in public safety.
We anticipate that these barriers will change in the near future.
Technological solutions for emergency response are receiving considerable political attention, including U.S. Congressional support through a 2019 Act to ``promote the use of the best available technology to enhance the effective and cost-efficient response to wildfires'' \cite{dingell2019conservation}.
Our interviews also exposed several key design challenges for using situated visualizations in the field, such as balancing complexity and simplicity and coupling data with context. 
However, 
there is little empirical or grounded guidance for crafting immersive visualizations.
Our interviews suggest a need to balance visualization simplicity with support for a broad set of tasks and data sources.
Future work should explore ways of providing necessary information in immersive visualizations
while retaining a relatively unobscured view of the physical environment.

\textit{\reviewnew{How can visualizations better consider hardware constraints?}}
Our work also demonstrated the importance of 
\reviewnew{new hardware platforms for field analysis and visualization. }
In our interviews, we learned of technologies already being incorporated in the field that would further bolster the proposed workflow and system architecture.
For example, earth scientists generally find mobile phone GPS insufficiently precise for many \reviewnew{measurements}; however, they 
\reviewnew{have workarounds that could inform more precise geospatial data integration.}
The localized connectivity provided by the high range portable router described by aerial firefighters would enable better teaming when using an integrated system such as FieldView.
Continued collaboration with field researchers about the technology already in use could yield \reviewnew{field-ready VA systems.} 
Visualization tools should consider how to best integrate these technologies to support richer and more robust analyses. \review{Further, though FieldView employs the Microsoft HoloLens, excessive weight, 
	reduced contrast in natural light, and challenges with spatially mapping large outdoor environments make existing HMDs 
	unreliable for real-world deployments. While we have begun exploring pass-through \reviewnew{solutions,}
	fieldwork offers novel design challenges for AR hardware.}

\section{Conclusion}
Visualizations have the power to transform data-oriented practices in field work by bringing data into the field. 
This integration can bridge spatial and temporal gaps by integrating data collection and analysis.
We explore how mobile and immersive visualization could transform field analysis practices by bridging these gaps through integrated data collection and visualization platforms.
We develop preliminary insight into how visualizations might support this domain in interviews with 10 experts from 5 different domains. Our results allowed us to elicit design guidelines and open questions for data analysis systems for fieldwork.
We instantiate this feedback in FieldView, an open-source prototype platform for the mobile and situated visualization of field-collected data.
This research provides preliminary steps \review{towards characterizing the novel design space of visualization for field analysis.}
\acknowledgments{\review{
		The authors thank the expert participants whose feedback drove this work. 
		This work was supported by NSF Award \#1764092 and an IGP grant from the University of Colorado Boulder.}}

\bibliographystyle{abbrv-doi}
\balance

\bibliography{fieldview}
\end{document}